\documentclass[twocolumn,secnumarabic,amssymb, nobibnotes, aps, prb, superscriptaddress]{revtex4-2}
\usepackage{lipsum}
\usepackage{titlesec}

\titlespacing\section{0pt}{12pt plus 4pt minus 4pt}{1pt plus 20pt minus 2pt}
\usepackage{xcolor}
\usepackage[colorlinks, allcolors=blue]{hyperref}
\usepackage{tabularx}
\usepackage{amsmath}
\usepackage{comment}
\usepackage{makecell}
\usepackage{chemformula}
\usepackage{afterpage}
\usepackage{siunitx}
\usepackage{gensymb}
\usepackage{placeins}
\usepackage{graphicx}
\usepackage{float}
\usepackage{booktabs}
\usepackage{multirow}
\usepackage{array}
\usepackage{setspace}
\graphicspath{{Figs/}}
\usepackage{siunitx}
\usepackage{hhline}
\usepackage{xfrac}
\usepackage{float,graphicx}
\usepackage{mathtools}
\usepackage{listings}
\usepackage{amssymb}
\usepackage{titlesec}
\usepackage{amsfonts}

\usepackage[version=4]{mhchem}

\usepackage{epstopdf}
\catcode`@11
\def\seceqaa{\@addtoreset{equation}{section}
\def\theequation{A\arabic{equation}}}
\def\seceqbb{\@addtoreset{equation}{section}
\def\theequation{B\arabic{equation}}}
\def\seceqcc{\@addtoreset{equation}{section}
\def\theequation{C\arabic{equation}}}
\def\seceqdd{\@addtoreset{equation}{section}
\def\theequation{D\arabic{equation}}}
\def\seceqee{\@addtoreset{equation}{section}
\def\theequation{E\arabic{equation}}}
\def\seceqff{\@addtoreset{equation}{section}
\def\theequation{F\arabic{equation}}}
\def\seceqgg{\@addtoreset{equation}{section}
\def\theequation{G\arabic{equation}}}
\def\seceqhh{\@addtoreset{equation}{section}
\def\theequation{H\arabic{equation}}}
\catcode`@11

\begin{document}

\title{\textbf{\textbf{\textbf{Unconventional topological Hall response and anisotropic magnetotransport properties of a helical magnet EuAuAs} } } }

\author{Anyesh Saraswati}
\affiliation{S. N. Bose National Centre for Basic Sciences, Salt Lake City, Kolkata-700106, India}

\author{Koyendrila Debnath}
\altaffiliation{These authors contributed equally to this work}
\affiliation{S. N. Bose National Centre for Basic Sciences, Salt Lake City, Kolkata-700106, India}

\author{Shubhankar Roy}
\altaffiliation{These authors contributed equally to this work}
\affiliation{Vidyasagar Metropolitan College, 39 Sankar Ghosh Lane, Kolkata 700006, India}

\author{Barun Ghosh}\email{bghosh@bose.res.in}
\affiliation{S. N. Bose National Centre for Basic Sciences, Salt Lake City, Kolkata-700106, India}

\author{Nitesh Kumar}\email{nitesh.kumar@bose.res.in}
\affiliation{S. N. Bose National Centre for Basic Sciences, Salt Lake City, Kolkata-700106, India}

\author{Prabhat Mandal}\email{prabhat.mandal@bose.res.in}
\affiliation{S. N. Bose National Centre for Basic Sciences, Salt Lake City, Kolkata-700106, India}

\begin{abstract}
Topological magnets with nontrivial spin texture have attracted considerable interest because they display a rich spectrum of emergent quantum phenomena. Here, we present a combined experimental and theoretical investigation of the magnetic and magnetotransport properties of EuAuAs, an antiferromagnet with Néel temperature ($T_\mathrm{N}$) $\sim$ 6 K. The temperature and magnetic field dependence of electrical resistivity and magnetization demonstrate that the charge transport in EuAuAs is strongly influenced by the spin configuration of local Eu moments. Below $T_\mathrm{N}$, both longitudinal magnetoresistance (LMR) and transverse magnetoresistance (TMR) are positive at low fields but large and negative at high fields. With increasing temperature, TMR becomes positive above 60 K, whereas LMR remains negative up to 100 K. The low-field positive LMR and TMR originate from weak antilocalization (WAL). The WAL contribution in TMR is well captured by the Hikami–Larkin–Nagaoka model, whereas the LMR data are described by a generalized Altshuler–Aronov framework. Moreover, we observe a giant topological Hall effect arising from the scalar spin chirality, which is further supported by the helical magnetic structure obtained from the ab-initio calculations. The observed anisotropy in longitudinal resistivity and magnetoresistance underscores the very nature of the Fermi surface of the EuAuAs, as confirmed by first-principles calculations. These results establish EuAuAs as a unique platform for exploring the interplay between electronic structure and noncoplanar spin texture in a centrosymmetric helical magnet.

\vspace{3mm}
\end{abstract}

\maketitle
\section{INTRODUCTION}
\vspace{3mm}
Lately, the domain of topological materials has expanded well beyond topological insulators to encompass a broad and rich family of topological semimetals -- Dirac, Weyl, nodal-line, and even higher-degeneracy phases. These semimetals are characterized by symmetry-protected band crossings in the bulk and emergence of distinctive boundary states \cite{hasan2010colloquium,qi2011topological,kumar2020topological},  which could lead to the formation of unconventional quasiparticles \cite{bradlyn2016beyond,wang2022axial,sarkar2023charge,sarkar2026kramers}.
Topological insulators exhibit a full bulk energy gap accompanied by symmetry-protected conducting surface or edge states. Topological semimetals, on the contrary, lack a complete bulk gap and the symmetry-protected band crossings
\cite{bradlyn2016beyond,armitage2018weyl,yan2017topological,   fang2016topological,roy2018magneto,lv2017observation,ma2018three, Singha2017} directly underpin several hallmark phenomena such as Fermi-arc surface states, backscattering suppression, spin–momentum locking, etc. \cite{barman2020symmetry,bernevig2022progress}. The non-trivial spin texture further enriches this landscape. When incorporated into band structure calculations, the breaking of time-reversal symmetry can generate electronic bands with nontrivial global invariants and host exotic quasiparticles. Experimentally, the presence of such quasiparticles is reflected through a variety of unconventional electrical and thermoelectric responses, such as anomalous Hall effects (AHE) \cite{nagaosa2010anomalous,singha2019magnetotransport,liu2018giant}, anomalous Nernst effects \cite{xiao2006berry,ikhlas2017large}, chiral anomaly-driven negative magnetoresistance (NMR) \cite{huang2015observation,balduini2024intrinsic}, and topological Hall effects (THE). Net scalar spin chirality from noncoplanar spin textures can generate THE signals  \cite{lee2020probing,kurumaji2019skyrmion,moya2022incommensurate,nabi2021giant}, arising from the real space Berry curvature (BC) without spin orbit coupling (SOC).

Weak antilocalization (WAL) is a quantum interference phenomenon observed in electronic transport at low temperature, particularly in topological materials with SOC, often manifested as a characteristic cusp in the magnetic field dependence of resistivity \cite{lu2014weak,laha2021topological,malick2022weak,hou2015transition}. In massless Dirac fermionic systems, a $\pi$ Berry phase results in destructive interference. Hence, carrier backscattering is suppressed, thereby enhancing conductivity and leading to the WAL effect due to strong carrier coherence.

Eu-based pnictide compounds with chemical formula Eu\textit{TX} (\textit{T} = Au, Ag, Cu; \textit{X }= pnictogens) have recently attracted considerable attention due to the complex and diverse nature of their physical properties  \cite{jin2021multiple,yin2025magnetism,ram2024magnetotransport,roy2024chiral,chi2024electronic,takahashi2023superconductivity,lipika2026complex,yin2025magnetism,wang2023structure,may2023coupling,wang2023structure}. These materials exhibit a rich variety of magnetic ground states, ranging from ferromagnetic (FM) to antiferromagnetic (AFM) ordering to more intricate helical spin configurations. In addition, several members of this family are proposed to host different topological phases of matter and exhibit various intriguing properties. 
For instance, EuCuP exhibits pronounced anisotropy in magnetotransport properties ‐- with first‐principle calculations linking this behavior to intrinsic BC effects in its FM state  \cite{wang2023anisotropic}. Similarly, EuCuSb shows frustrated quasi-FM behavior, and its spin modulation is coupled to charge carriers, giving rise to anomalous and topological Hall responses \cite{wang2025large}. Both EuAgAs and EuCuAs are AFM Weyl semimetals and exhibit unusual magnetotransport properties with large THE \cite{laha2021topological,roychowdhury2023interplay}. EuAuBi exhibits canted AFM order along with THE and signature of superconductivity below 3 K \cite{lipika2026complex,singh2026intrinsic,takahashi2023superconductivity}. More recently, apart from THE, EuAuSb single crystals were found to exhibit both NMR and WAL up to $\sim$ 100 K, and quantum oscillations up to 70 K, demonstrating the presence of topological transport in Eu\textit{TX} magnets \cite{roy2024chiral}.

These striking observations naturally motivated us to investigate the closely related compound EuAuAs to know whether it also hosts a topological phase and exhibits large THE comparable to its sister compounds EuAgAs and EuCuAs.  As already mentioned, EuAuSb hosts a topological semimetallic phase, while EuAuBi exhibits superconductivity. This sharp contrast, even within the same Au-based series, naturally positions the sister compound EuAuAs as a compelling platform for in-depth study.

In this work, we present a combined experimental and theoretical investigation of the electronic, magnetic, and transport properties of EuAuAs single crystals. We have noticed that based on first-principals calculations, Mallick \textit{et al.}  predicted a nodal-line band structure in EuAuAs \cite{malick2022electronic}. However, their experimental results do not provide any conclusive evidence supporting the proposed topological nature of the electronic bands. Furthermore, in contrast to their prediction of a collinear AFM ground state, our detailed theoretical calculations reveal that a helical spin structure is energetically more stable. Consequently, we uncover several unconventional signatures in transport properties: a large, two-component THE featuring a sharp low-field peak tied to a metamagnetic transition  (MMT)   and a broad intermediate-field hump arising from scalar spin chirality, along with a strong Fermi surface (FS) anisotropy-driven magnetotransport and WAL effects.

\vspace{3mm}
\section{EXPERIMENTAL AND THEORETICAL METHODS}

\vspace{3mm}

\subsection{Single crystal growth and characterization}

Single crystals of EuAuAs were synthesized by the flux method with bismuth as flux. Pieces of high-purity elements (Alfa Aesar),  europium (Eu, 99.9 $\%$), gold (Au, 99.9 $\%$), arsenic (As, 99.99 $\%$), and bismuth (Bi, 99.99 $\%$)  were mixed in a molar stoichiometric ratio 1:1:1:10, and the mixture was loaded into an alumina crucible. The crucible was sealed in a quartz tube under partial argon pressure with quartz wool as a filter. The ampoule was placed in a muffle furnace, heated to 1100$^\circ$C at a rate of 30$^\circ$C/h and held at this temperature for 12 h. The furnace was then slowly cooled to 600$^\circ$C at a rate of 3$^\circ$C/h, and the bismuth flux was decanted using a centrifuge. Single crystals of EuAuAs of different sizes were obtained. Before any magnetic and transport measurements, the crystals were polished to remove any trace of bismuth flux on the surface. X-ray diffraction (XRD) was carried out on single crystals and powdered single crystals using an x-ray diffractometer (Smart Lab, Rigaku) equipped with Cu-K$_{\alpha}$ radiation. For powder XRD, we randomly picked up small single crystals and crushed them into fine powder without polishing. The obtained XRD data were refined using the FullProf software. To analyze the elemental composition of the obtained crystals, we have performed the emissive dispersive spectroscopy (EDS) on a field emission scanning electron microscope (Quanta 250 FEG). Magnetic measurements were performed using the vibrating sample magnetometer (VSM) in a physical properties measurement system (PPMS, Dynacool, Quantum Design). The resistivity measurements under an external magnetic field ($B$) were performed using the electrical transport option (ETO) of the PPMS. A symmetrization method was employed to eliminate any small erroneous transverse contribution due to misalignment of voltage leads, as defined by the formula $\rho_{xx}$(\textit{B}) = [$\rho_{xx}$(+\textit{B}) + $\rho_{xx}$(-\textit{B})]/2, from the longitudinal resistivity. Similarly, an antisymmetrization approach was implemented to remove the erroneous longitudinal contribution in the case of Hall resistivity measurements, using the relation $\rho_{yx}$(\textit{B}) = [$\rho_{yx}$(+\textit{B}) - $\rho_{yx}$(-\textit{B})]/2. Several independent magnetotransport measurements were performed, and the results are reproducible within experimental error.

\vspace{3mm}

\begin{figure}[t]
\centering
\includegraphics[width=0.5\textwidth]{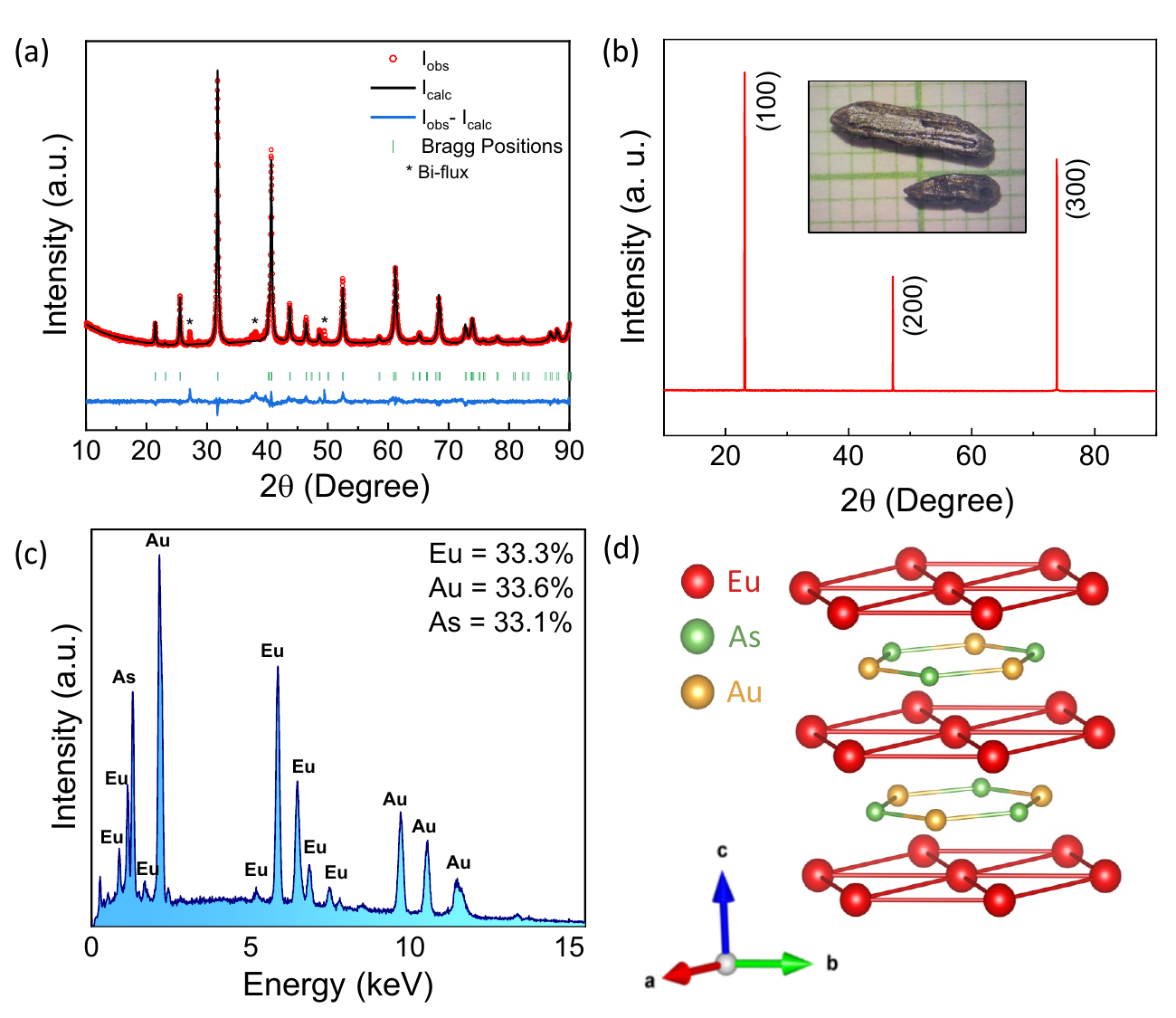}
%\centering
\caption{(a) Powder XRD of EuAuAs taken on finely crushed samples. The inset shows the picture of as-grown single crystals. (b) Single crystal XRD of one of the crystals along the (\textit{l}00) direction. (c) EDS Spectra of single crystalline EuAuAs. (d) Schematic of the structure of a unit cell of EuAuAs.}
\label{fig1}
\label{fig1}
\end{figure}

\vspace{3mm}

\subsection{Computational Details}

\vspace{3mm}

Our first-principles calculations are based on density functional theory (DFT) using the plane-wave projected augmented wave (PAW) as implemented in Vienna ab initio Simulation Package (VASP) \cite{blochl1994projector, kresse1999ultrasoft, kresse1996efficiency, kresse1996efficient}. We have adopted a generalized gradient approximation (GGA) parameterized by Perdew, Burke, and Ernzerhof (PBE) for the exchange-correlation functional \cite{perdew1996generalized}. The plane-wave energy cut-off was set to 500 eV, and a 13$\times$13$\times$4 Gamma-centered \textit{k}-mesh was used for Brillouin zone (BZ) sampling. We have included SOC for all our theoretical analyses. To account for the correlations between the \textit{f}-electrons at the magnetic Eu sites, we have adopted the GGA+\textit{U} method with the value of $\textit{U}=5$ eV \cite{dudarev1998electron}. We performed ionic relaxations of the internal degrees of freedom until the residual forces on each atom were less than $10^{-5}$ eV/\AA{}. We have considered both in-plane and out-of-plane AFM spin configurations and calculated the total energies and electronic structure for six magnetic configurations, (i) AFM-A \cite{roy2024chiral} using the $1\times 1\times 1$ conventional unit cell, (ii) AFM-z using $1\times 1\times 1$ conventional unit cell, (iii) DP-AFM \cite{roy2024chiral} using a $1\times1\times2$ supercell, (iv) h-AFM-90, where the direction of Eu spin rotates by $90^{\circ}$ between the successive layers, using a $1\times1\times2$ supercell, (v) h-AFM-60, a helical configuration, where the direction of Eu spin rotates by $60^{\circ}$ between successive layers and (vi) h-AFM-120, where the spin moments of Eu atoms rotate by $120^{\circ}$ between successive layers. A supercell of $1\times1\times3$ has been considered for both h-AFM configurations (h-AFM-60 and h-AFM-120). Although the magnetic periodicity in h-AFM-120 is defined by three successive Eu moments, the presence of intervening non-magnetic Au and As layers between Eu planes makes it necessary to construct a $1\times1\times3$ crystallographic supercell to represent one complete helical magnetic period within periodic boundary conditions.
\vspace{3mm}

\section{RESULTS AND DISCUSSIONS}
\vspace{3mm}

\subsection{Structural and elemental analysis}

 The powder XRD was performed on a few crushed crystals at room temperature, as depicted in Fig. \ref{fig1}a. It confirms the absence of impurity phases, except for a few small peaks corresponding to Bi (marked with $\star$).  This is because the surface of the unpolished crystals contains a tiny amount of Bi flux. The flat surface of a relatively large crystal (inset of Fig. \ref{fig1}b) was placed parallel to the sample holder, and the obtained XRD resulted in only (\textit{l}00) peaks (Fig. \ref{fig1}b), suggesting that the \textit{a}-axis aligns normal to the surface. The representative EDS spectra of a crystal, as shown in Fig. \ref{fig1}c, reveal a near-perfect stoichiometry, Eu:Au:As = 1:1.01:0.99, and this ratio is almost the same for different crystals in the batch. EuAuAs crystallizes into a ZrSiBe-type centrosymmetric hexagonal structure with space group $P6_3/mmc$ (No. 194) and the extracted lattice parameters are  \textit{a} = \textit{b} = 4.438 Å and \textit{c} = 8.28 Å, as determined by Rietveld refinement (Table S1 of supplementary information SI \cite{supply}). It consists of repeating layers of Eu-(Au+As)-Eu as shown in Fig. \ref{fig1}d. Six triangles consisting of corner-sharing Eu atoms coalesce into a hexagon. Each Eu-layer is separated by hexagons comprising corner-sharing Au and As atoms, arranged alternately.

 \subsection{Magnetic properties}
 
\begin{figure}
\centering
\includegraphics[width=0.5\textwidth]{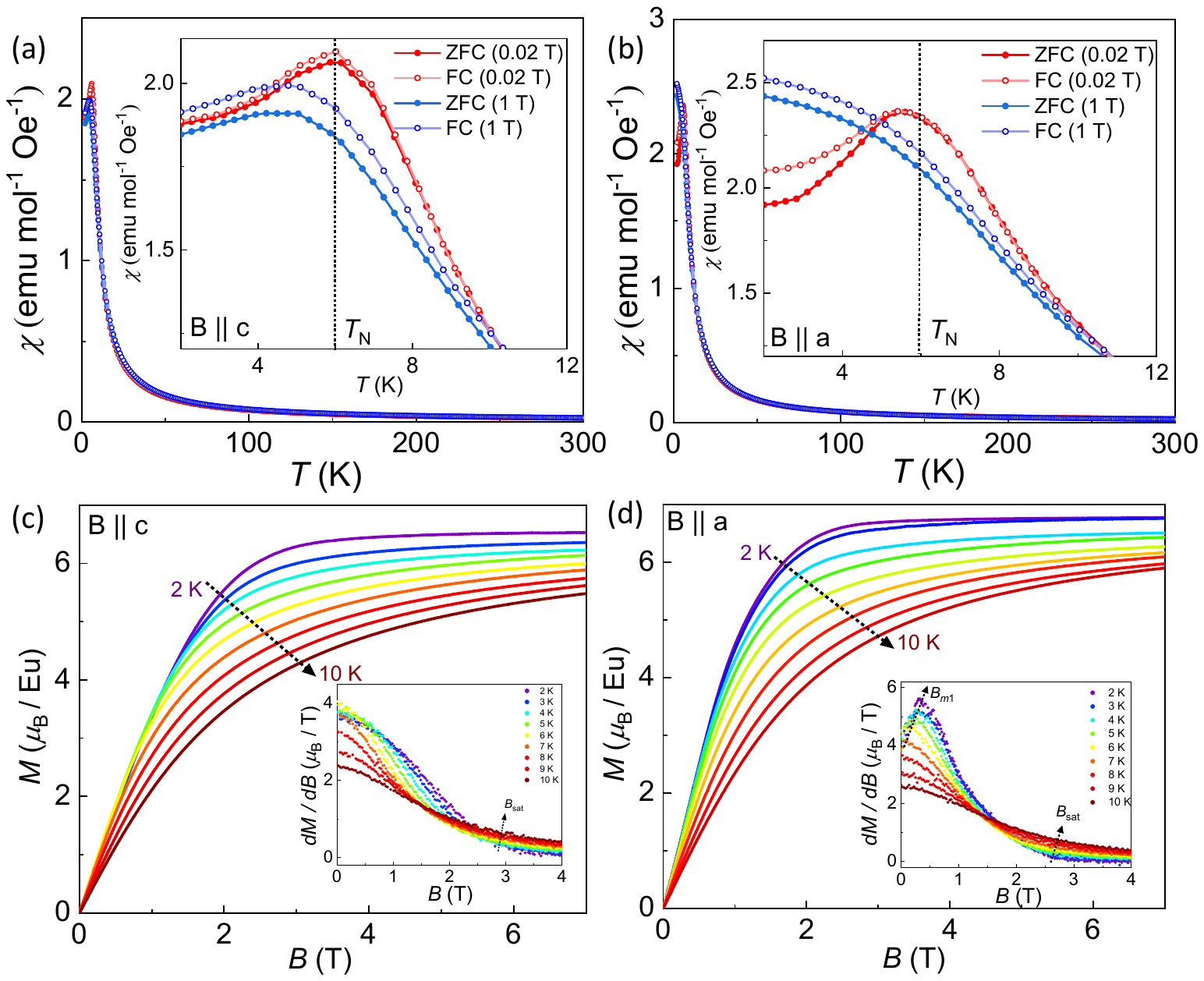}
\caption{Temperature dependence of magnetic susceptibility for EuAuAs with magnetic field parallel to (a) {$c$}-axis and (b) {$a$}-axis. (c) and (d) Magnetic field dependence of magnetization for EuAuAs at a few representative temperatures with applied field parallel to {$c$}- and {$a$}-axes, respectively.}
\label{fig2}
\end{figure}

The temperature dependence of dc magnetic susceptibility, $\chi$(=$M/B$), of the EuAuAs crystal was measured with the field \textit{B} applied parallel to both \textit{c} ($B \parallel $ $c$) and \textit{a} ($B \parallel $ $a$) axes configurations. Both the field-cooled (FC) and zero-field-cooled (ZFC) magnetic data at fields of 0.02 and 1 T are depicted in Figs. \ref{fig2}a and  \ref{fig2}b. With decreasing temperature,  the $\chi (T)$ along both the \textit{c} and \textit{a} axes exhibit a sharp peak at  $T_\mathrm{N}$ = 6 K in both FC and ZFC cycles, where EuAuAs undergoes a transition from a paramagnetic (PM) to an AFM state (Inset of Figs. \ref{fig2}a and  \ref{fig2}b). Along with the transition, the insets also clearly point out the anisotropic nature of the magnetic properties.  The  inverse of susceptibility data (Fig. S1 of SI \cite{supply}) were fitted in the temperature range of 50 to 300 K  employing the modified Curie-Weiss (MCW)  equation, 
\begin{equation}
\chi (T) = \chi_0 + \frac{C}{T - \theta_\mathrm{W}},
\end{equation}
where $\chi_0$ is the temperature-independent magnetic susceptibility and \textit{C} and $\theta_{\mathrm{W}}$ are the Curie constant and Weiss temperature, respectively. The fitting yields  $\theta_{\mathrm{W}}$ $\sim$ 5.4 and 4.6 K for the $B \parallel $ $a$ and $B \parallel $ $c$ directions, respectively.  It may be noted that $\theta_\mathrm{W}$ is positive in both directions and its value is close to $T_\mathrm{N}$, which suggests that the dominant interaction is FM in nature, consistent with the earlier report \cite{malick2022electronic}. The value of effective magnetic moment along both crystallographic directions, determined using the relation $\mu_{eff}$ = $\sqrt{8C}$ $\mu_\mathrm{B}$, is found to be 7.8 $\mu_\mathrm{B}$.  This is in excellent agreement with the theoretical value for the free Eu$^{2+}$ ion (7.94 $\mu_\mathrm{B}$).

\begin{figure*}
\includegraphics[width=18cm]{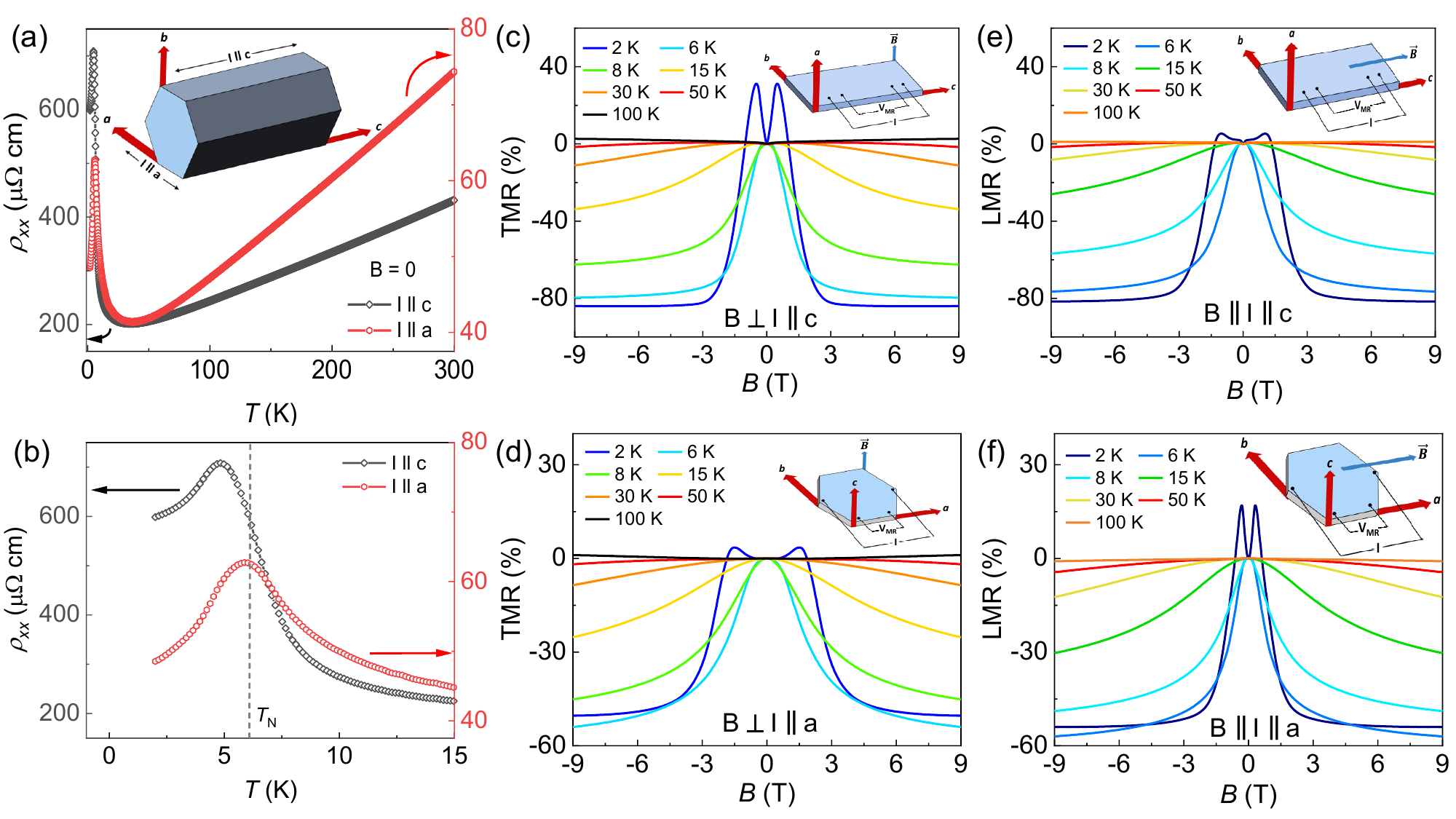}
\caption{(a) The temperature dependence of the longitudinal resistivity ($\rho_{xx}$) measured along different directions for the EuAuAs single crystal at zero field condition. The inset shows a schematic of the measurement configuration. (b) Zoomed in view of the temperature-dependent $\rho_{xx}$ shown up to 5 K, $T_\mathrm{N}$ representing the magnetic transition temperature. The magnetic field dependence of TMR for (c) $B\perp I \parallel c$, (d) $B\perp I \parallel a$, and the corresponding field dependence of LMR for (e) $B\parallel I \parallel c$, (f) $B\parallel I \parallel a$ at a few representative temperatures up to 100 K.}
\label{fig4}
\end{figure*}

The magnetization isotherms as a function of magnetic field $M(B)$  across the magnetic transition are illustrated in Figs. \ref{fig2}c and \ref{fig2}d for the $B \parallel $ $c$ and $B \parallel $ $a$ configurations, respectively. For $B \parallel$ $a$, the magnetization curve at 2 K initially increases rapidly up to $\sim$ 2 T and then tends to saturate with further increase in \textit{B}. The same trend is observed for $B \parallel$ $c$, but the saturation-like behavior of $M$ starts at a slightly higher field $\sim$ 3 T. This suggests that the $a$ axis is the easy axis of magnetization in the EuAuAs crystal. The observed values of saturation magnetization are 6.76 and 6.82 $\mu_\mathrm{B}$ for $B \parallel $ $c$ and $B \parallel $ $a$, respectively, which are approximately equal to the expected spin-only moment of a free Eu$^{2+}$ ion. In addition, the $M$($B$) curves below 6 K exhibit an upward curvature accompanied by a subtle hysteresis below a critical field $B_{m1}$ ($\sim 0.34$ T at 2 K) for $B \parallel $ $a$ configuration, which is a signature of a MMT (inset of Fig. \ref{the}c). This critical field is defined by the peak position in $dM/dB$ vs $B$ curve, which moves progressively towards lower field with increasing temperature and disappears at 6 K. This weak but distinct transition is further highlighted in the inset of Fig. \ref{fig2}d. However, no such field-induced transition is observed for the $B \parallel$ $c$ configuration (inset of Fig. \ref{fig2}c).

\subsection{Resistivity and magnetoresistance}

Figure \ref{fig4}a displays the temperature dependence of longitudinal resistivity $\rho_{xx}(T)$ measured with the current along the \textit{c} and \textit{a} directions. In both cases, $\rho_{xx}$ decreases with decreasing temperature down to 35 K, exhibiting a typical metallic behavior ($d\rho_{xx}/dT>0$). $\rho_{xx}$ is approximately linear in $T$ above 70 K. The observed metallic behavior in EuAuAs is consistent with our electronic band structure calculations, where the density of states of As contributes significantly at the Fermi energy ($E_\mathrm{F}$) (Fig. \ref{fig9}b). Below 35 K, the  $\rho_{xx}(T)$  exhibits a sharp upturn, which is attributed to the scattering of conduction electrons due to spin fluctuations as the system approaches the magnetic transition. The sharp drop in resistivity below 6 K is attributed to the suppression of spin-dependent scattering as a result of the onset of long-range ordering of Eu$^{2+}$ moments. At 2 K, $\rho_{xx}$ for $I\parallel c$ is about 12 times higher than that for $I\parallel a$, indicating a significant anisotropy (Fig. \ref{fig4}b). Similar behavior was also observed in EuCuP \cite{wang2023anisotropic} and EuZnGe \cite{kurumaji2022anisotropic}. Indeed, our DFT calculations clearly show that the FS of EuAuAs is highly anisotropic in nature  (see Fig. \ref{fig9}d). This is further supported by pronounced anisotropy in carrier mobility as well as a strong dependence of resistivity on the direction of the applied magnetic field (discussed in later sections).
  
To gain further insight, we have measured the magnetoresistance (\textit{MR}) of the sample in both the $B \perp I\parallel c$  and $B \perp I\parallel a$ configurations. \textit{MR} is defined as $\frac{\rho_{xx}(B) - \rho_{xx}(0)}{\rho_{xx}(0)}\times100 $, where $\rho_{xx}(B)$ and $\rho_{xx}(0)$ correspond to the resistivities measured at magnetic field $B$ and zero fields, respectively. The $\rho_{xx}(B)$  measured at various temperatures with the field perpendicular to the current flow (TMR configuration) is shown in Figs. \ref{fig4}c  and \ref{fig4}d. The measurement schematic is depicted in the insets of Fig. \ref{fig4}c (\ref{fig4}d), where the field was applied along \textit{a} (\textit{c}) direction, and the current was applied along \textit{c} (\textit{a}) direction. The corresponding isotherms are depicted in the main panel of Fig. \ref{fig4}c and \ref{fig4}d. For the configuration $B \perp I\parallel c$, \textit{MR} increases gradually with increasing field, exhibiting a cusp-like feature with a maximum value of $30 \%$ at $\sim $ 1 T. With further increase in field, \textit{MR} decreases rapidly, becomes negative, and reaches a value $\sim$ 84 $\%$ at 9 T and 2 K. The cusp-like feature weakens with increasing temperature and disappears above $T_\mathrm{N}$, and \textit{MR} is negative over the whole range of field with no sign of field-induced magnetic transitions up to 9 T. A similar behavior is observed for the configuration $B\perp I \parallel a$. However, in this case, the cusp appears at a slightly lower field, and the maximum value of MR is $\sim$ 10 $\%$. \textit{MR} reaches a maximum (negative) value  $\sim$ 54 $\%$ at 2 K and 9 T. 

In a similar fashion, the $\rho_{xx}(B)$ was also measured with the field and current parallel to each other in both \textit{a} and \textit{c} directions (LMR configuration) and displayed in Figs. \ref{fig4}e and \ref{fig4}f along with the measurement schematics in their respective insets. In this case, the LMR exhibits a sharp cusp at low field, and reaches a maximum value 82 $\%$ at 9 T and 2 K for $I\parallel B\parallel c$. Similarly, a maximum value of 54 $\%$ was observed for the $I\parallel B \parallel a$ configuration under the same conditions. It may be mentioned that we observe a significant anisotropy in TMR as well as in LMR for both configurations. This reflects the anisotropic nature of the electronic structure of the system, and is consistent with the cylindrical nature of FS (See Fig. \ref{fig9}d). Generally, in a quasi-two-dimensional electronic system, where charge carriers are confined within the plane, a magnetic field applied parallel to the plane does not affect the orbital motion as strongly as that for the field perpendicular to the plane of the crystal, resulting in an anisotropy in \textit{MR}. For $I \parallel a$, the LMR is larger than the TMR. On the other hand, for $I \parallel c$, an opposite behavior is observed up to 40~K, where the TMR is larger; above 40~K, the trend reverses. A few representative plots depicting this anisotropic behavior in LMR and TMR are shown in Fig. S2 of the SI for a better visualization \cite{supply}.

\begin{figure*}
\includegraphics[width=18cm]{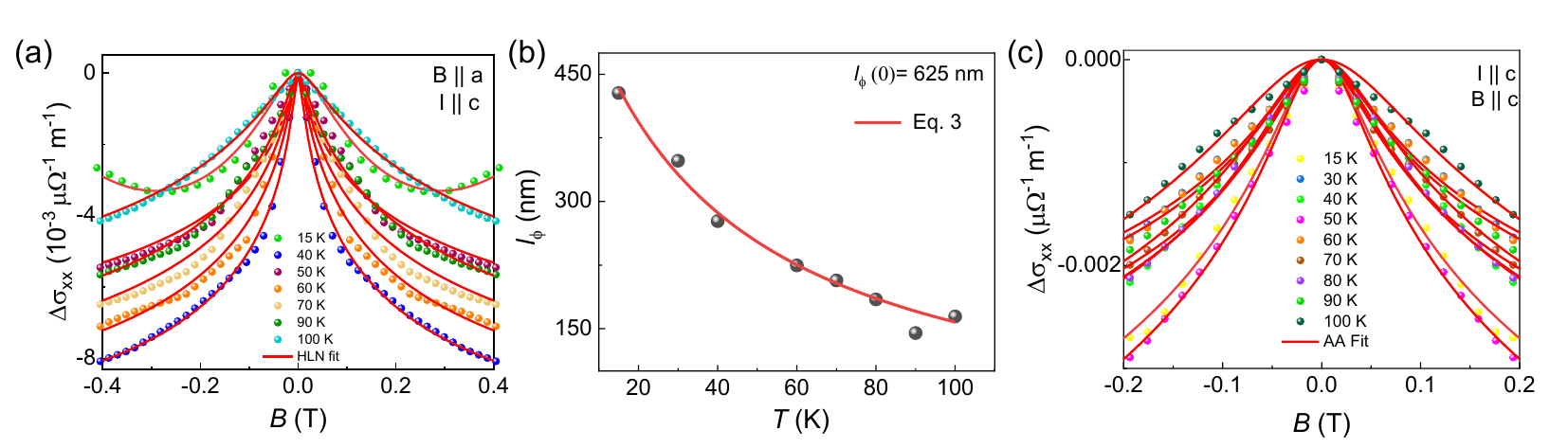}
\caption{(a) Field dependence of transverse magnetoconductivity at different temperatures in the low-field region. The solid red lines show the fit to Eq. (2) (HLN model). (b) $l_{\phi}$ as a function of temperature with fitting of Eq. (3). (c) Longitudinal magnetoconductivity as a function of magnetic field. The solid lines show the fit to Eq. (4) (AA model).}
\label{fig5}
\end{figure*}

The observed cusp-like feature in TMR and LMR is attributed to the WAL effect of the charge carriers. Although this effect is most pronounced in two-dimensional (2D)  electron systems, a robust WAL signature has also been detected in a variety of three-dimensional topological insulators and semimetals  \cite{laha2021topological,malick2022weak,xu2014weak,sasmal2020magnetotransport,hou2015transition,shrestha2017extremely}. Fundamentally, WAL arises when coherent electron waves traversing time-reversed trajectories interfere destructively, thereby reducing the likelihood of electrons retracing closed paths and becoming localized. This mechanism produces a distinct low-field cusp in MR, reflecting the enhanced phase coherence and strong SOC in these materials \cite{bergmann1982weak}. Consequently, WAL also serves as a powerful probe of surface or low-dimensional transport channels across both strictly two-dimensional platforms and topologically nontrivial bulk compounds. To analyze this, we calculate the magnetoconductance (MC) by employing the formula $\sigma_{xx} (B) = \rho_{xx}/(\rho_{xx}^2 + \rho_{yx}^2)$,  $\rho_{yx}$
being the  Hall resistivity. We now employ the modified Hikami-Larkin-Nagaoka  (HLN) equation to fit the TMR data. Although this model was developed for 2D systems, several topological systems also satisfy the equation very well \cite{malick2022weak,sasmal2020magnetotransport,saraswati2025single,hou2015transition,roy2024chiral,chamorro2019dirac}. To accommodate the SOC and elastic scattering contributions, we also include a $B^2$ term. The modified HLN equation for TMR is now defined in terms of transverse MC difference [$\Delta\sigma_{xx} (B) = \sigma_{xx} (B) - \sigma_{xx} (0)$] as \cite{hikami1980spin},
\begin{equation}
\Delta\sigma_{xx} =  -\frac{\alpha e^2}{\pi h}\left[\psi\left(\frac{1}{2} + \frac{\hbar}{4el^2_{\phi}B}\right) - \mathrm{ln\left(\frac{\hbar}{4el^2_{\phi}B}\right)}\right] + vB^2,
\end{equation}
where $\alpha$ represents the number of channels involved in the conduction process, $\psi$ is the digamma function, $l_\phi$ is the phase coherence path of the charge carrier, and $v$ is the strength of other scattering contributions. In order to avoid complications arising from the carrier scattering due to the magnetically ordered spins, the fitting was done well above $T_\mathrm{N}$. The transverse MC fits reasonably well with the modified HLN equation in the low-field regime, depicted in Fig. \ref{fig5}a, with the $B^2$ term further improving the quality of fit. The fit reveals $\alpha\sim$ 10$^5$, which is significantly larger than that expected for a 2D system ($\alpha$ = 1/2 for each coherent conducting channel), and similar large value for was also observed earlier in several topological systems \cite{laha2021topological,malick2022weak,xu2014weak,sasmal2020magnetotransport,hou2015transition,roy2024chiral}, further signifying the WAL originating from the strong SOC of the 3D bulk state with multiple conduction channels. $l_\phi$ decays with increasing temperature, signifying the suppression of the WAL effect due to increased frequency of inelastic scattering at higher temperatures. The dependence of $l_\phi$ on temperature can be interpreted by assuming both electron-electron (e-e) and electron-phonon (e-p) scattering terms in the form of the relation \cite{shrestha2017extremely},
\begin{equation}
    \frac{1}{l_\phi^2} =  \frac{1}{l_\phi^2 (0)} + AT + BT^2,
\end{equation}
where $l_\phi (0)$ is the phase coherence length at 0 K, and \textit{A }and \textit{B} are the coefficients of e-e and e-p scattering, respectively. Fig. \ref{fig5}b demonstrates the temperature dependence of  $l_\phi$. The fit to data (solid line) yields $l_\phi (0)$ = 625 nm, \textit{A} = 1.49 x 10$^{-7}$ nm$^{-2}$ K$^{-1}$, and \textit{B} = 2.27 x 10$^{-9}$ nm$^{-2}$ K$^{-2}$.

The LMR data were fitted in the low-field region with the generalised Altshuler-Aronov (AA) model \cite{lin2013parallel}, 
\begin{equation}
  \Delta\sigma_{xx}  = -\frac{\alpha e^2}{\pi h}\ln\left[1 + k\left(\frac{edl_e}{\hbar}\right)^2B^2\right ],
\end{equation}
where \textit{d} is the sample thickness, and $l_e$ is the mean free path of the charge carrier. The value of \textit{k} is determined by the strength of the disorder present in the system. The fitting was, again, performed at temperatures well above the magnetic transition. On the basis of the disorder strength, the scattering process can be segregated into several regimes. For instance, in a highly disordered system, $l_e$ could be significantly smaller than the sample dimension ($l_e<<d$). In such a scenario, \textit{k} is equal to 1/3, and it is known as the AA regime. However, for a very clean metallic thin film, $l_e$ could exceed the sample thickness ($l_e>>d$) at low temperatures, and the charge scattering mechanism in this weakly disordered regime can be described by the Dugaev-Khmelnytskyi model with \textit{k} = 1/6. As we have performed the experiments on a bulk single crystal with a thickness of \textit{d} $\sim$ 0.37 mm, it is reasonable to assume $l_e<<d$. The experimental longitudinal MC data reasonably fit with the AA equation up to temperatures as high as 100 K and have been displayed in Fig. \ref{fig5}c. The value of $l_e\sim$ 1 nm, which is much smaller than \textit{d}, indicates that the electron transport occurs in the quantum diffusive regime. The order of $\alpha \sim 10^5$ is consistent with that obtained from the earlier HLN fit of the TMR data. Closer inspection of Fig. \ref{fig4} and Fig. S2 (See SI \cite{supply}) shows that TMR is larger than LMR below 30 K for current along the high symmetry axis ($c$ axis). However, this trend reverses above 30 K. This excess magnetoconductivity in LMR at high temperature well above $T_\mathrm{N}$ may be attributed to the chiral-anomaly-induced charge conduction.

\begin{figure}
\includegraphics[width=1.0\linewidth]{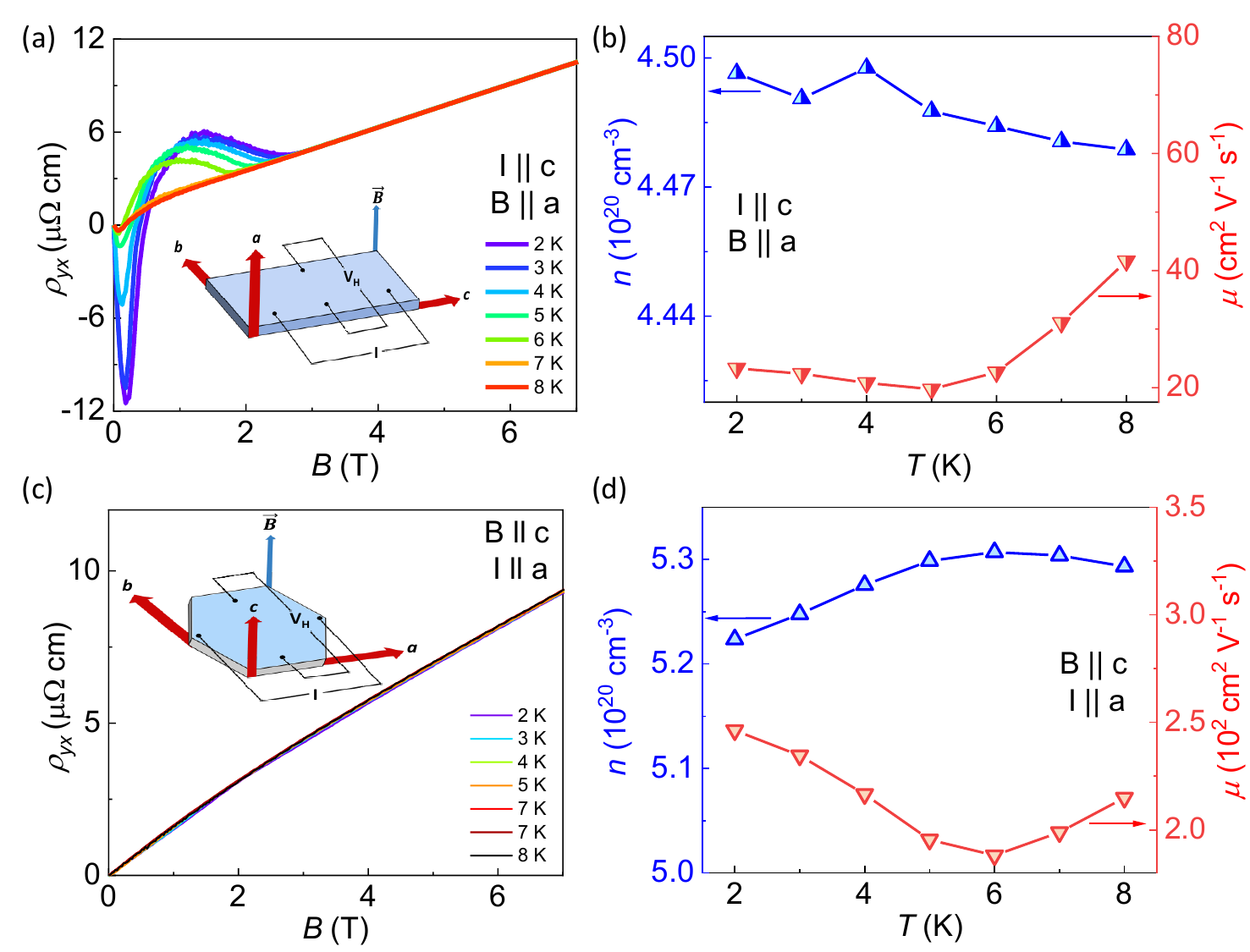}
\caption{(a) Magnetic field dependence of Hall resistivity $\rho_{yx}$ of EuAuAs at different temperatures, and (b) temperature variation of carrier density and mobility for the configuration  $B\parallel a$ and $I\parallel c$. (c) Magnetic field dependence of $\rho_{yx}$ of EuAuAs at different temperatures, and (d) temperature variation of carrier density and mobility for the configuration  $B\parallel c$ and $I\parallel a$. }
\label{fig6}
\end{figure}

\begin{figure*}
\centering
\includegraphics[width=1.0\linewidth]{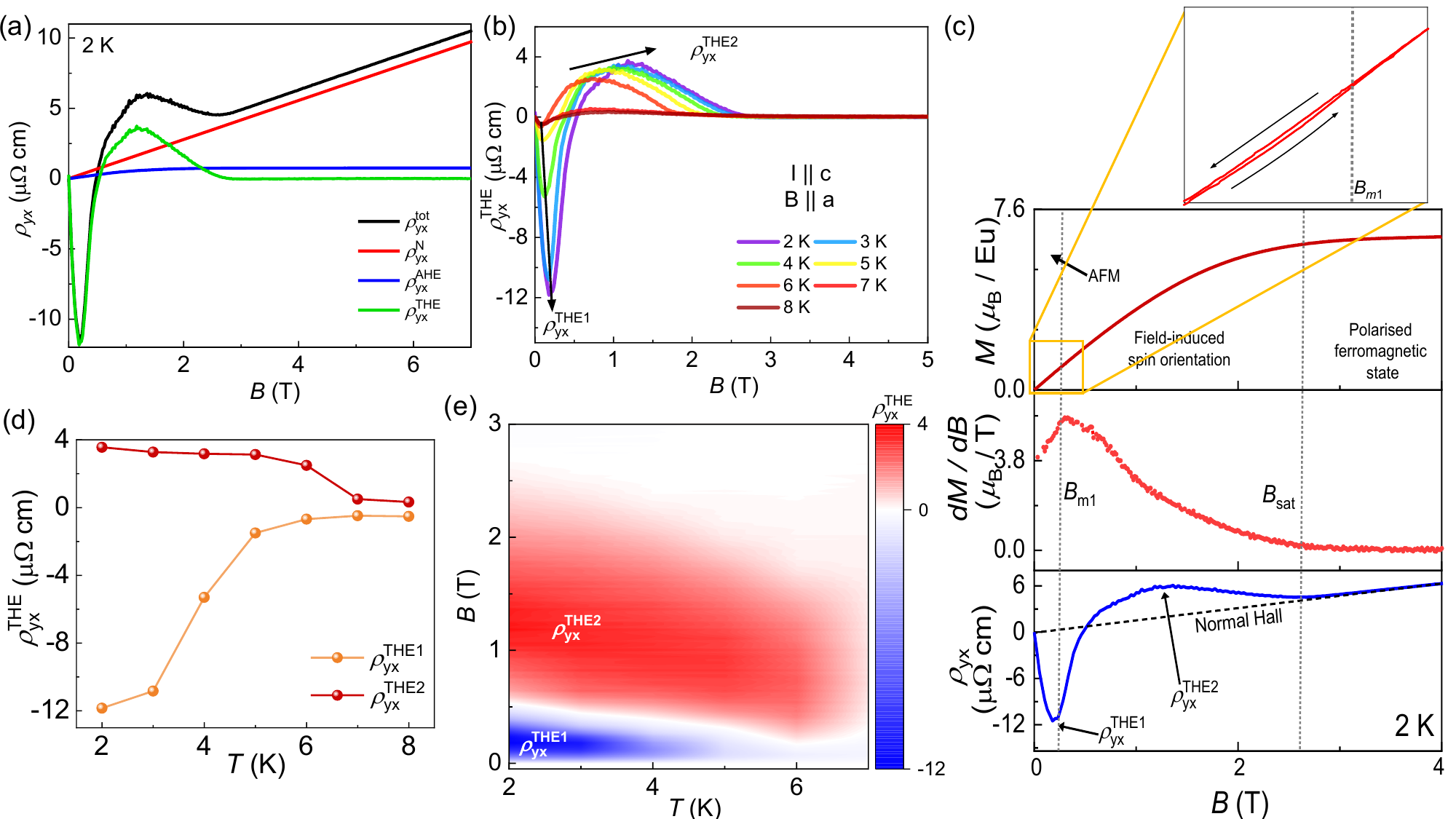}
\caption{(a) Representation of the different components of the Hall resistivity represented at 2 K for $B\parallel a$ and $I\parallel c$. (b) Magnetic field dependence of topological Hall resistivity at different temperatures. (c) Vertical panel displaying the magnetic field dependence of \textit{M}, \textit{dM/dB}, $\rho_{\mathrm{yx}}$, and \textit{d$\rho_{\mathrm {yx}}$/dB} at 2 K. Inset highlights the low-field hysteresis of the magnetisation curve. (d) The temperature dependence of the maxima of $\rho^{\mathrm{THE1}}_{\mathrm{yx}}$ and  $\rho^{\mathrm{THE2}}_{\mathrm{yx}}$ of EuAuAs.(e) Contour plot displaying the variation of the topological Hall resistivity with \textit{B} and \textit{T}. }
\label{the}
\end{figure*}

\subsection{Hall resistivity}

The Hall resistivity $\rho_{yx}$ was measured as a function of magnetic field for two configurations: $I \parallel c$ (Fig.~\ref{fig6}a) and $I \parallel a$ (Fig.~\ref{fig6}c). At high-field, as the magnetization-dependent contributions are either frozen or effectively saturated, only the ordinary Hall component remains, and $\rho_{yx}$ is linear in $B$. Within the single-band approximation, the carrier density $n$ and mobility $\mu$ are given by $n = 1/(R_0 e)$ and $\mu = 1/(n e \rho_{xx})$, where $e$ is the electron charge and $R_0$ is the ordinary Hall coefficient obtained from the high-field linear fit. The positive slope of $\rho_{yx}$ versus $B$ line  (Figs.~\ref{fig6}a and \ref{fig6}c) indicates hole-dominated conduction. The carrier concentration is of the order of $10^{20}~\text{cm}^{-3}$, confirming the semimetallic nature of EuAuAs. At 2~K, we obtain $\mu = 22~\text{cm}^2\text{V}^{-1}\text{s}^{-1}$ for $I \parallel c$, whereas, for  $I \parallel a$, the value of $\mu$ = $247~\text{cm}^2\text{V}^{-1}\text{s}^{-1}$. This pronounced mobility anisotropy is due to the cylindrical nature of the FS of EuAuAs, with its principal axis along $k_z$. Such a geometry leads to enhanced scattering for transport along the out-of-plane direction, thereby producing markedly different mobilities along two crystallographic directions. These anisotropic carrier dynamics underpin the observed anisotropy in both electrical resistivity and MR data.

For the configuration with current along \textit{a}-axis, $\rho_{yx}$ shows no additional features, which confirms the absence of any anomalous contribution. For the configuration with the current along the $c$-axis, as we begin to move towards the low-field regime, a hump-like feature starts to appear below $T_\mathrm{N}$ in the mid-field region of $B = 0.5-2.5$ T, the amplitude of which diminishes in the polarized FM region above 2.5 T  (Fig. \ref{fig6}a). At first glance, it may appear that the hump originates from the anomalous contribution. However, as the $\rho_{yx}$ does not follow the magnetization isotherm precisely (Fig \ref{fig2}a), the hump also has a different origin, and we attribute it to the THE. When electrons pass through non‑coplanar spin arrangements, they pick up a real‑space Berry phase from the scalar spin chirality of the local moments, which acts like an emergent magnetic field and produces an extra transverse voltage, on top of the ordinary and anomalous Hall signals. \cite{bruno2004topological,gobel2025topological,he2022topological}. This emergent field can be described by the Berry phase acquired by electrons adiabatically following the orientation of three localized spins, mathematically proportional to the scalar spin chirality $\textbf{S}_i. (\textbf{S}_j \times \textbf{S}_k)$, which acts like a magnetic flux density in real space \cite{chen2025topological}. Thus, an additional contribution arises in $\rho_{yx}(B)$ from THE, and can be expressed as,
\begin{equation}
 \rho_{yx} (B) = \rho_{yx} ^N + \rho_{yx} ^{AHE} + \rho_{yx} ^{THE}  = R_0B + R_s\mu_0M + \rho_{yx}^{THE}
\end{equation}
where $\rho_{yx} ^N$ and  $\rho_{yx} ^{AHE}$ are the ordinary and anomalous contributions to the total Hall resistivity, with $R_0$ and $R_s$ being the ordinary and anomalous Hall coefficients, respectively. The THE can be determined from the total Hall resistivity data by the relation  $\rho_{yx}^{THE} (B)$ = $\rho_{yx}$ - ($\rho_{yx} ^N$ + $\rho_{yx} ^{AHE}$). The normal and anomalous Hall components can be scaled experimentally in the high-field regime presented in Fig. \ref{the}a. The extracted THE is plotted in Fig. \ref{the}b, where the hump is localized in the aforementioned field range, and the peak shifts to a lower value of \textit{B} on increasing temperature and progressively becoming broader. The peak exhibits a maximum value of $\sim$ 3.5 $\mu\Omega$ cm at 2 K, a substantially large value among known topological systems \cite{gong2021large,huang2022plateau,li2019large,rout2019field,li2020large,wang2021field}. This effect fades away at higher \textit{B}, attributed to the disappearance of the non-coplanar spin structure due to complete polarization of spins along the field direction. 

Below $T_\mathrm{N}$,  we observe an additional sharp peak-like anomaly at around 0.3 T, whose amplitude also progressively diminishes with increasing temperature. This feature is highly unusual, exceeding the magnitude of the conventional THE hump and differing markedly from typical topological Hall signals reported in other systems. To elucidate its origin, we have performed the field derivatives of the magnetization isotherms at 2 K as shown in Fig.~\ref{the}c. The derivative $dM/dB$ exhibits a maximum at $B_{m1}$, which directly links the sharp feature to the field-induced MMT. 
Such a transition in frustrated systems can proceed via nucleation and motion of domain walls, which act as chiral scattering centers for conduction electrons and could produce an anomalous Hall signal in the Hall component. The $\rho_{yx}^{\mathrm{THE1}}$ amplitude drops by ~87\% at 5~K, indicating strong coupling to the magnetization.
The maximum values of the THE at both regimes were plotted as a function of temperature (Fig. \ref{the}d), along with a contour plot in Fig.~\ref{the}e. They further reveal the coexistence of two distinct signatures in the low-temperature region (below $T_\mathrm{N}$): a sharp anomaly for $0<B<B_{m1}$ confined to a narrow low-field ($B<0.3$~T) associated with spin orientation and domain-wall dynamics, reflecting a delicate field-sensitive state, whereas the other part spans a much broader field ($B_{m1}<B<B_{\mathrm{sat}}$) and temperature range, corresponding to a conventional THE arising from finite scalar spin chirality in a stabilized noncoplanar spin configuration.

It may also be noted that the double-peak Hall resistivity feature observed in EuAuAs bears a striking resemblance to the Hall response predicted for a double‑\textbf{Q} (2\textbf{Q}) magnetic structure \cite{ohgata2026large}, where a 2\textbf{Q} spin texture, even without net scalar spin chirality, may generate a large AHE that exhibits a nonmonotonic dependence on the out‑of‑plane magnetization. This theoretical behavior is perfectly captured by our double peak feature in our experimental data (see Fig.~\ref{the}b). Moreover, the predicted enhancement of the $\rho_{yx}$ in the 2\textbf{Q} state, several times larger than that of the FM phase, is consistent with the large magnitude of the THE measured in EuAuAs. While our first‑principles calculations establish a helical AFM ground state for EuAuAs, there remains a strong possibility that a field‑induced multi‑\textbf{Q} spin texture contributes significantly to the Hall signal, with the features arising from multi‑\textbf{Q}–induced BC hot spots. This interpretation is further supported by the fact that Eu$^{2+}$ carries zero orbital angular momentum, resulting in the nearly isotropic magnetic behavior observed in Figs.~\ref{fig2}a and~\ref{fig2}b. Such isotropic exchange allows the spins to freely modulate under an applied field and readily allows the formation of a multi‑\textbf{Q} structure.

\vspace{3mm}

\begin{figure}
\includegraphics[width=1.0\linewidth]{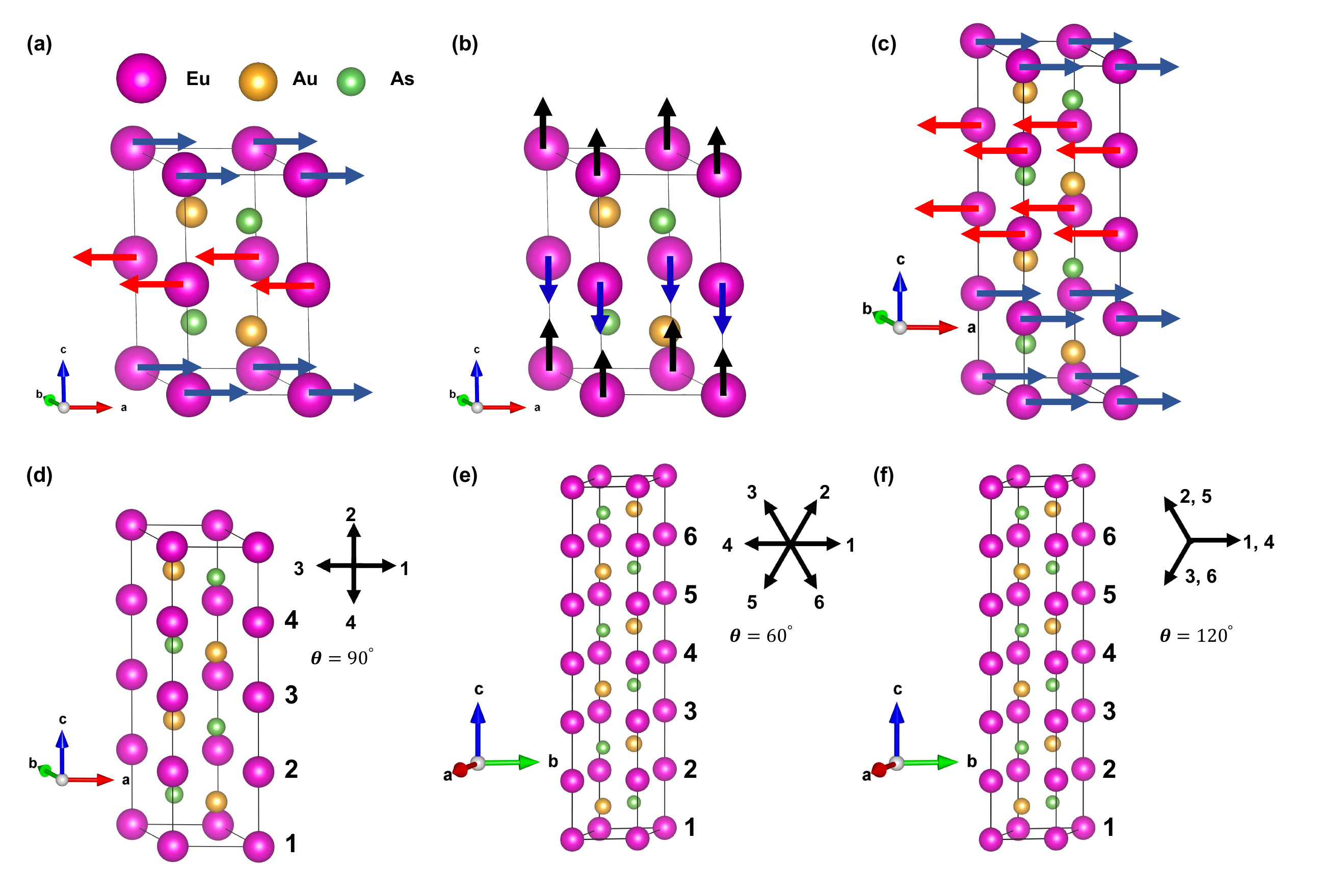}
  \caption{Crystal structures of EuAuAs with (a) AFM-A, (b) AFM-z, both using a ($1\times 1\times 1$) conventional unit cell, (c) DP-AFM shown using a ($1\times 1\times 2$) supercell, and (d) h-AFM-($90^\circ$) magnetic configurations shown using a ($1\times 1\times 2$) supercell, and (e) h-AFM-($60^\circ$) and (f) h-AFM-($120^\circ$) configurations, shown using a ($1\times 1\times 3$) supercell of the hexagonal unit cell. The arrows indicate the directions of the magnetic moments on the Eu atoms. The side panels in (d)–(f) show top views of the in-plane magnetic-moment components in the Eu layers labelled by the corresponding numbers.}
\label{fig8}
\end{figure}

\section{First principles calculations}
\vspace{3mm}
As EuAuAs crystallizes in a hexagonal structure with space group $P6_3/mmc$, it preserves inversion symmetry. The optimized lattice constants of the 1$\times$1$\times$1 unit cell EuAuAs are $a=b=4.49$ {\AA} and $c=8.17$ {\AA}. These are in good agreement with previous reports \cite{malick2022electronic, pottgen2000equiatomic}, and our Rietveld refinement \cite{supply}. The conventional unit cell contains six atoms, where the Eu layer is sandwiched between two Au-As layers stacked along the \textit{z}-direction. As shown in Fig.~\ref{fig8}, multiple magnetic configurations of EuAuAs were constructed by varying the relative orientation of the localized Eu moments between successive Eu layers along the $c$ axis. In the AFM-A configuration, the Eu moments are aligned within the basal plane, while adjacent Eu layers are antiferromagnetically coupled, resulting in a collinear two-layer magnetic periodicity. In the DP-AFM configuration, the Eu moments also lie in the $ab$-plane, but the magnetic order consists of ferromagnetically aligned Eu bilayers that are antiferromagnetically coupled to neighboring bilayers, giving rise to an enlarged four-layer magnetic periodicity along $c$ axis. In the AFM-z configuration, the Eu moments are oriented parallel to the $c$ axis, while maintaining AFM coupling between neighboring Eu layers, thus preserving the same two-layer periodicity with out-of-plane spin polarization. In addition to these collinear magnetic states, we consider three commensurate helical AFM (h-AFM) configurations: h-AFM-$90^\circ$, h-AFM-$60^\circ$, and h-AFM-$120^\circ$. In the h-AFM-$90^\circ$ state, the Eu moments rotate within the basal plane by $90^\circ$ between successive Eu layers, forming a four-sublattice helical spin texture with a magnetic period described by a $1\times1\times2$ supercell. In case of h-AFM$-60^\circ$ configuration, the spin direction rotates by $60^\circ$ between adjacent Eu layers, completing a full $360^\circ$ rotation over six Eu layers, which requires a $1\times1\times3$ supercell. Similarly, in the case of h-AFM$-120^\circ$, the Eu moments undergo a $120^\circ$ rotation between successive layers. Importantly, in this case, the magnetic orientation repeats periodically every fourth Eu layers; however, the non-magnetic Au–As layers enforce a magnetic unit cell that is a $1\times1\times3$ supercell of the original non-magnetic unit cell. The relative ground state energies and the magnetic moments per Eu atom for different magnetic configurations are listed in Table 1. For Hubbard $U = 5$ eV, the h-AFM-$60^\circ$ and h-AFM-$120^\circ$ are nearly degenerate in energy, the energy difference being 0.1 meV, which is also sensitive to the details of the computational parameters. Upon increasing the Hubbard $U$, the relative stability of the two magnetic configurations changes, and for $U = 6$ eV the h-AFM-120 state becomes energetically more favorable than the h-AFM-60 state. We thus conclude that the stability of the phases is sensitive to the value of Hubbard $U$ chosen. A recent single-crystal neutron diffraction study on the sister compound EuAuSb~\cite{sears2025euausb} has revealed an incommensurate helical magnetic order in which successive FM layers rotate in-plane by approximately \(120^\circ\). Motivated by these observations and the DFT results, we present the electronic structure of the h-AFM-120$^\circ$ configuration in the main text. The electronic structure for other magnetic configurations is provided in Fig. S3 of SI for completeness \cite{supply}. For all magnetic configurations, the calculated local magnetic moment at the Eu site is approximately $6.9~\mu_\mathrm{B}$, consistent with the localized $4f^7$ electronic configuration and the divalent Eu$^{2+}$ oxidation state.

\begin{table}[t]
\centering
\caption{Relative energies and local magnetic moments for different magnetic configurations of EuAuAs calculated with $U = 5$ eV.}
\vspace{2mm}
\renewcommand{\arraystretch}{1.2}
\setlength{\tabcolsep}{0pt}   % remove fixed spacing to let \extracolsep work
\begin{tabular*}{\columnwidth}{l @{\extracolsep{\fill}} c c}
\hline
Configuration &  \makecell{Relative energy in \\meV per Eu atom} & \makecell{Local moment on \\Eu atom ($\mu_\mathrm{B}$)} \\
\hline
AFM-A & 0.73 & 6.90 \\
AFM-z & 1.46 & 6.90 \\
DP-AFM & 2.83 & 6.89 \\
h-AFM-$90^\circ$ & 3.63 & 6.89 \\
h-AFM-$120^\circ$ & 0.0 & 6.90 \\
h-AFM-$60^\circ$ & -0.05 & 6.90 \\
\hline
\end{tabular*}
\label{tab:mag_configs}
\end{table}

\begin{figure}
\includegraphics[width=1.0\linewidth]{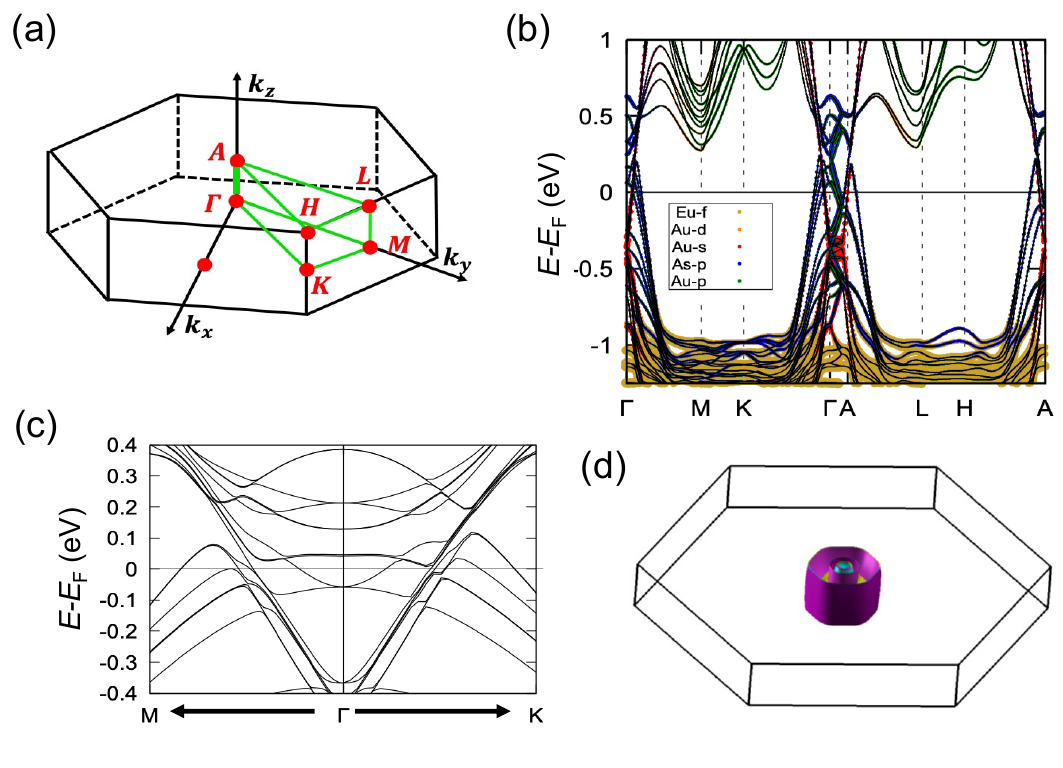}
      \caption{(a) Hexagonal Brillouin zone (BZ) of EuAuAs with various high-symmetry points marked. (b) Orbital resolved band structure along the high-symmetry directions of EuAuAs for h-AFM-$120^\circ$ magnetic configuration.  Near the Fermi level, the orbital contributions of the bands mostly arise from the Au s and Au p and As p orbitals. The localized Eu \textit{f} bands are also evident. (c) Zoomed-in bands along the M--$\Gamma$--K direction showing both doubly degenerate and nondegenerate states near the Fermi level. (d) Calculated Fermi surface of EuAuAs in the helical antiferromagnetic state, h-AFM$-120^\circ$, showing the anisotropic character of the electronic structure.}
\label{fig9}
\end{figure}

The electronic band structure of h-AFM-$120^\circ$ magnetic configuration along the high-symmetry path (Fig.~\ref{fig9}a)  of the Brillouin zone (BZ) is shown in Fig.~\ref{fig9}b. The overall band structure is metallic, with noticeable band crossing between the valence and conduction bands around the $\Gamma$ point. The orbital resolved band structure (Fig.~\ref{fig9}b) suggests that near $E_\mathrm{F}$, the orbital contributions of the bands mostly arise from the Au-$s$ and Au-$p$ and As-$p$ orbitals. The Eu-$4f$ states appear as nearly dispersionless bands located below $E_\mathrm{F}$, reflecting their strongly localized Eu$^{2+}$ character, while the Eu-$5d$ states contribute mainly to the higher-energy conduction bands. Along the M--$\Gamma$ direction (Fig.~\ref{fig9}c), six bands originating predominantly from As-$p$ and Au-$d$ orbitals are observed, among which two sets of bands remain almost doubly degenerate with a few meV SOC-induced band splitting, while the remaining two bands are nondegenerate.

The FS of the h-AFM-$120^\circ$ configuration of EuAuAs reveals a strongly anisotropic and quasi-two-dimensional electronic structure, intimately connected to the layered hexagonal crystal lattice and underlying magnetic order (Fig.~\ref{fig9}d). In the in-plane direction, the Fermi contour consists of a nearly circular pocket with a clear hexagonal warping centered around the $\Gamma$ point, consistent with the intrinsic hexagonal symmetry of the lattice and indicative of an almost isotropic electronic dispersion within the basal plane. In contrast, along the out-of-plane direction, the FS evolves into an extended, weakly warped cylindrical pocket with only a small modulation along $k_z$. Such a cylindrical FS topology reflects the dominant orbital hybridization within the $ab$-plane and comparatively weaker interlayer hopping along the $c$ axis, giving rise to pronounced anisotropy in the electronic transport between the in-plane and out-of-plane directions.

\section{CONCLUSION}
\vspace{3mm}
In this work, we have studied the magnetic and magnetotransport properties of high-quality single crystals of EuAuAs, an itinerant AFM belonging to the family of pnictide-based topological semimetals. Our results reveal that the magnetic and transport properties of the system are strongly influenced by the underlying spin configurations. Both longitudinal resistivity and magnetoresistance show pronounced anisotropy. Resistivity along the $c$ axis is an order of magnitude higher than that along the $a$ axis at 2 K. This behavior is directly linked to the cylindrical nature of the FS, which features dominant orbital hybridization within the $ab$-plane and weaker hopping along the $c$ axis. At low magnetic fields, EuAuAs exhibits the WAL effect, signifying SOC and carrier coherence. 
The material exhibits a massive THE response, reaching a maximum of $\sim$ 3.5 $\mu\Omega$ cm at 2K, which is substantial compared to other known topological systems.
THE response is uniquely composed of two distinct parts. The first contribution, $\rho_{yx}^{THE1}$, appears as a sharp, unconventional, low-field peak in the range 0–0.3 T, which coincides with an MMT, indicating a critical spin dynamics and domain-wall motion that act as effective scattering centers for charge carriers. The second contribution, $\rho_{yx}^{THE2}$, is manifested as a broad hump in the range 0.5 and 2.5 T, which represents a conventional THE arising from finite scalar spin chirality associated with a helical magnetic structure. 
Contrary to earlier predictions of a collinear AFM state, detailed first-principles calculations and experimental signatures establish that EuAuAs is a helical magnet. We believe that the present results will lead to further investigation and help to construct a theoretical model for in-depth understanding of this class of materials.

\vspace{3mm}

\section*{ACKNOWLEDGEMENTS}
BG acknowledges the support received from the Prime Minister’s Early Career Research Grant (PM-ECRG) from the Anusandhan National Research Foundation (ANRF), File Number ANRF/ECRG/2024/003677/PMS, and also benefited from the PARAM-Rudra computational facility at SNBNCBS. SR acknowledges the support received from  UGC-DAE-CSR Mumbai, India (Ref: CRS/21-22/03/365). NK acknowledges funding under Advanced Research Grant (ARG) Programme from ANRF, File Number ANRF/ARG/2025/004788/PS and Max Planck Society for funding under the Max Planck-India partner group project. PM acknowledges the support from the Raja Ramanna Chair (RRC) scheme of the Department of Atomic Energy, India. This research project made use of the instrumentation facility provided by the Technical Research Centre (TRC) at the S. N. Bose National Centre for Basic Sciences, under the Department of Science and Technology, Government of India.
\vspace{3mm}

\bibliography{reference}

\end{document}